\documentclass[english,twocolumn,aps,prl, superscriptaddress,showpacs, amsmath]{revtex4-1}

\usepackage{color,xcolor}

\usepackage{natbib}
\usepackage[T1]{fontenc}
\usepackage{units}
\usepackage{amssymb,amsmath}
\usepackage{graphicx}
\usepackage{esint}
\usepackage{bm}
\usepackage{ulem}
\usepackage[colorlinks=true, citecolor=blue]{hyperref}
\setcitestyle{journalcolor=blue}

\begin{document}

\title{Enhancement of Superconductivity upon reduction of carrier density in proximitized graphene}

\author{Gopi Nath Daptary}
\author{Udit Khanna}
\author{Eyal Walach}
\author{Arnab Roy}
\author{Efrat Shimshoni}
\author{Aviad Frydman}

\affiliation{Department of Physics, Jack and Pearl Resnick Institute and the Institute of Nanotechnology and Advanced Materials, Bar-Ilan University, Ramat-Gan 52900, Israel}

\date{\today}

\begin{abstract}
The superconducting transition temperature ($T_c$) of a single layer graphene coupled to an Indium oxide (InO) film, 
a low carrier-density superconductor, is found to increase with  \textit{decreasing} carrier density 
and is largest close to the average charge neutrality point in graphene. Such an effect is very surprising in 
conventional BCS superconductors. We study this phenomenon both experimentally and theoretically. Our analysis suggests that the InO film induces random electron and hole-doped puddles in the graphene. 
The Josephson effect across these regions of opposite polarity enhances the Josephson coupling between the superconducting clusters in InO,
along with the overall $T_{c}$ of the bilayer heterostructure.
This enhancement is most effective when the chemical potential of the system is tuned between the charge neutrality 
  points of the electron and hole-doped regions. 
\end{abstract}

\maketitle


Low carrier-density superconductivity has been a topic of great interest in condensed matter research since its 
discovery in SrTiO$_3$~\cite{PhysRevLett.12.474}. In conventional BCS superconductors,  
the critical temperature, $T_c$, is known to increase with increasing carrier density ($n$)~\cite{tinkham2004introduction}.
Contrarily, experiments on a number of exotic low density superconductors, such as Li-intercalated layered 
nitrides~\cite{PhysRevLett.97.107001, PhysRevB.98.064512}, underdoped La$_{2-x}$Sr$_x$CuO$_4$~\cite{PhysRevLett.101.057005} 
etc., detected an enhancement of $T_c$ with decreasing $n$. These results were interpreted as evidence for a 
non-BCS mechanism of electronic pairing, such as electron-electron (rather than electron-phonon) 
interactions~\cite{PhysRevB.47.5202, PhysRevLett.114.077001}. To date there is no known mechanism for enhancement of 
$T_c$ upon reducing $n$ for a BCS superconductor. In this paper, we present results of a conventional superconducting system 
in which $T_c$ is largest close to a charge neutrality point (CNP) for which $n$ can be extremely small.

Two-dimensional superconductors, in which the chemical potential can be modulated by gate voltage ($V_g$) are an ideal system for 
approaching the ultra low carrier-density regime. Graphene~\cite{Geim} is unique in this sense since the low energy dispersion is 
linear with momentum, i.e., the conduction and valence band touch at discrete points (Dirac points) resulting in a gapless 
semiconductor~\cite{RevModPhys.83.407}. Hence $n$ can be tuned through the 
CNP and may, in principle, be as small as desired.
In this paper we show that coupling graphene to a highly disordered, low-density superconductor gives rise to unique situation as the superconducting islands induce hole-doped regions within graphene, thus generating two CNPs (discussed later) in place of the global Dirac point for the non-proximitized graphene. This leads to a unique situation where superconductivity is enhanced with decreasing $n$ and is strongest close to the average CNP. We present a model to explain this extraordinary result based on the Josephson effect between regions of opposite polarity within the graphene. We show that the Josephson coupling between different superconducting regions is maximal when the system is tuned approximately half-way between the 
charge neutrality points of the electron and hole-doped regions. This occurs close to the global CNP of the graphene layer in the heterostructure.

\begin{figure}[t]
\begin{center}
\includegraphics[width=0.48\textwidth]{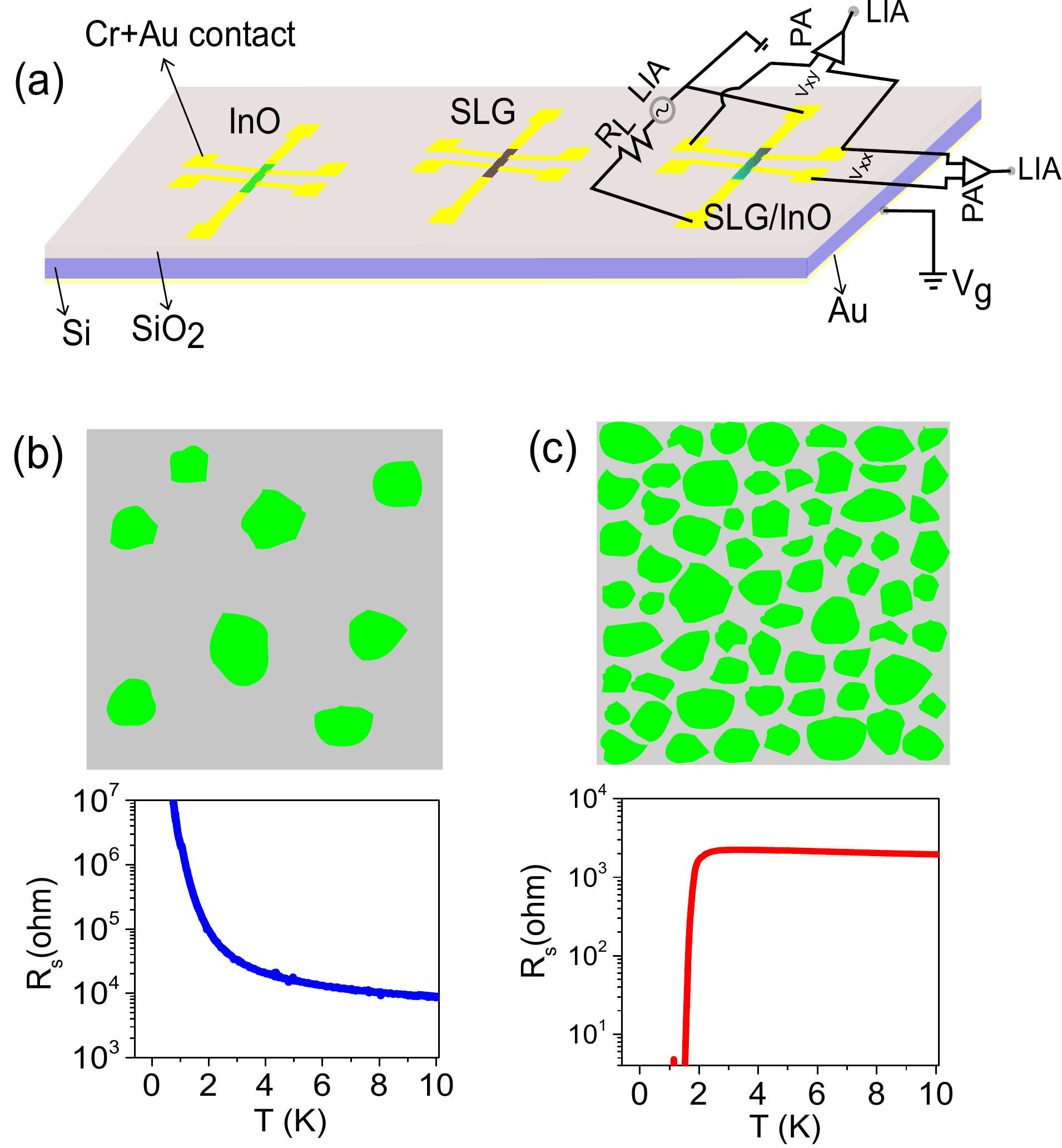}
\small{\caption{(a) A schematic diagram of the devices (From left: InO, SLG and an InO/SLG heterostructure). The longitudinal and transverse voltages are measured  by a lock-in amplifier (SR 830) after amplification of the signals by a low noise preamplifier (PA-SR552). The carrier density is modulated by the back gate voltage $V_g$ applied to the contact at the bottom of the Si. (b) and (c) Sketches of the superconducting islands and the resistance versus temperature curves of samples I and S respectively.
\label{sketch}}}
\vspace{-0.8cm}
\end{center}
\end{figure}

The experiments were performed on heterostructures of single layer graphene (SLG) and thin amorphous indium oxide (InO). We use CVD grown SLG sheets transferred onto 285 nm SiO$_2$ on top of a Si wafer as a 2D material. The sample was patterned into a Hall bar geometry by standard e-beam lithography and contacted to Cr/Au leads (5 nm/30 nm). It was then covered by a 30 nm thick InO film via a second lithography step. For reference, we prepared similar  geometries of bare graphene and bare InO [see Fig.~\ref{sketch}(a)]. The channel length and width of the sample are 150 $\mu$m and 50 $\mu$m respectively. The carrier density of the graphene device was modulated by changing the gate voltage applied to the back side of Si wafer. The device structure along with the electrical connections are shown in Fig.~\ref{sketch}(a). Measurements were performed in a wet He-3 system at temperatures down to 0.3 K.

InO  is a low density superconductor where $n$ can be controlled between $\sim 10^{19} - 10^{20} \text{ cm}^{-3}$ 
by changing the O$_2$ partial pressure during film deposition~\cite{ovadyahu1986some}. For large $n$, the critical temperature, $T_C$ can reach $\sim 3.5K$ and the coherence length $\xi$ is $30-50 nm$ \cite{johansson2005nanowire,poran2011disorder}. Decreasing $n$ causes the InO film 
to undergo a transition from a superconducting state to an insulating state. Nevertheless it has been shown that in both phases, 
the film includes emergent superconducting puddles, with sizes of a few $\mu m$, embedded in an insulating matrix~\cite{kowal1994disorder,kowal2008scale,bouadim2011single}. 
Indeed, comparable finite energy gap, $\Delta$, and vortex motion were measured in both phases~\cite{sacepe2011localization, spathis2008nernst,
poran2011disorder,sherman2012measurement,kopnov2012little,roy2018quantum}. The difference between a superconducting film and 
an insulating one lies in the global superfluid density which depends on the Josephson coupling between superconducting puddles \cite{RevModPhys.91.011002}. 
This is illustrated schematically in Fig.~\ref{sketch} panels b and c which show the resistance versus temperature curves of two 
InO films: one insulating, denoted as sample I [panel (b)] and one superconducting denoted as sample S [panel (c)] together with 
sketches of the inherent superconducting granularity. In the insulating phase the superconducting islands are sparse and decoupled, so that 
superconductivity is present only locally, while in the superconducting phase Josephson coupling percolates across the sample 
and global superconductivity is achieved.

\begin{figure}[t]
\begin{center}
\includegraphics[width=0.48\textwidth]{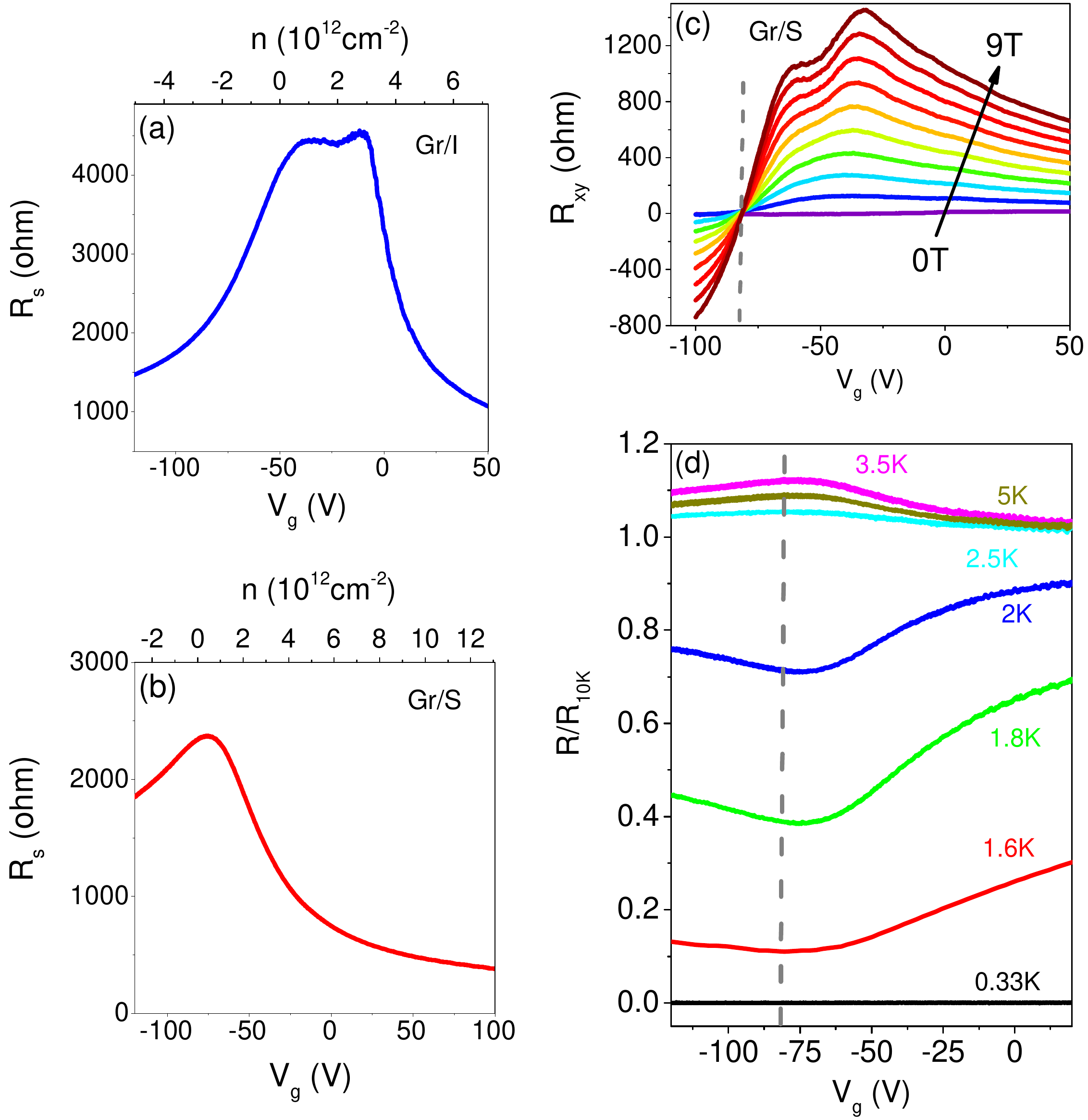}
  \small{\caption{(a) and (b) Sheet resistance, $R_s$, as a function of $V_g$ of sample Gr/I and sample Gr/S respectively at zero magnetic field. The measurements were performed at $T=1.7$ K for Gr/I and at $T=5$ K ($T>T_c$) for  Gr/S. (c) Hall resistance, $R_{xy}$, as a function of $V_g$ at different magnetic fields ($B=0-9$T in steps of $1$T) at $T=1.7$K of sample Gr/S. Note that the charge neutrality point is at a gate voltage of $V_d=-81.5$V. (d) Sheet resistance,  normalized by the resistance at 10K, as a function of $V_g$ at different $T$ of sample Gr/S. The slight difference between the CNP extracted from the Hall measurement and that of the resistance peak is attributed to the disorder of the sample which leads to some spatial distribution of $n$. Note that the resistance reaches a maximum at $T=3.5K$. This is due to the non-monotonic nature of indium oxide film transport \cite{roy2018quantum}. 
\label{RVg}}}
\vspace{-0.8cm}
\end{center}
\end{figure}
In this paper we discuss the results from two of the samples, Gr/S and Gr/I, which are heterostructures of SLG and a thin InO layer in the superconducting or insulating phase respectively. A second superconducting sample (Gr/S2) showed similar results as shown in the supplementary material. In a previous work we presented results on sample Gr/I~\cite{daptary2020superconducting}. In such a system the rather sparse InO superconducting  puddles proximitize the underlying regions in the graphene sheet, at the same time hole-doping them relative to the remaining 
SLG. Hence, the system includes a second charge neutrality point in addition to the usual CNP of the overall electron doped graphene (DP$_e$)~\cite{daptary2020superconducting}. This point, dubbed the 
``hole Dirac point'' (DP$_h$), gives rise to an additional peak in the resistance versus gate voltage (R-$V_g$) curve as seen in Fig.~\ref{RVg}(a). 
Unlike most experiments of SLG coupled to a BCS superconductor, the low carrier density of InO (a few orders of magnitude smaller than conventional superconductors) makes it experimentally possible  to access both CNPs, i.e DP$_e$ and DP$_h$ in sample Gr/I. 
Our results also indicate that in samples for which the  InO film is closer to the superconducting  transition, the separation (in energy) between  DP$_e$ and DP$_h$ is larger~\cite{daptary2020superconducting},
thus making it experimentally difficult to probe both CNPs. Nevertheless, a large region around the midpoint between DP$_e$ and DP$_h$ is accessible. 

In the current work, we focus on Gr/S.  Figure~\ref{RVg}(b) shows that for  high temperatures $T$ significantly above $T_C$, one resistance peak is observed. Hall effect measurements [see Fig.~\ref{RVg}(c)] identify this resistance peak as the charge neutrality point of the system. 
Surprisingly, as the temperature is lowered  close and below $T_C$, the peak at the CNP turns into a dip which becomes sharper with decreasing $T$, until a sufficiently low temperature at which the sample becomes superconducting in the entire $V_g$ regime [see Fig.~\ref{RVg}(d)]. This dip implies that superconductivity is \textit{strongest}  close to the CNP.


\begin{figure}[t]
\begin{center}
\includegraphics[width=0.48\textwidth]{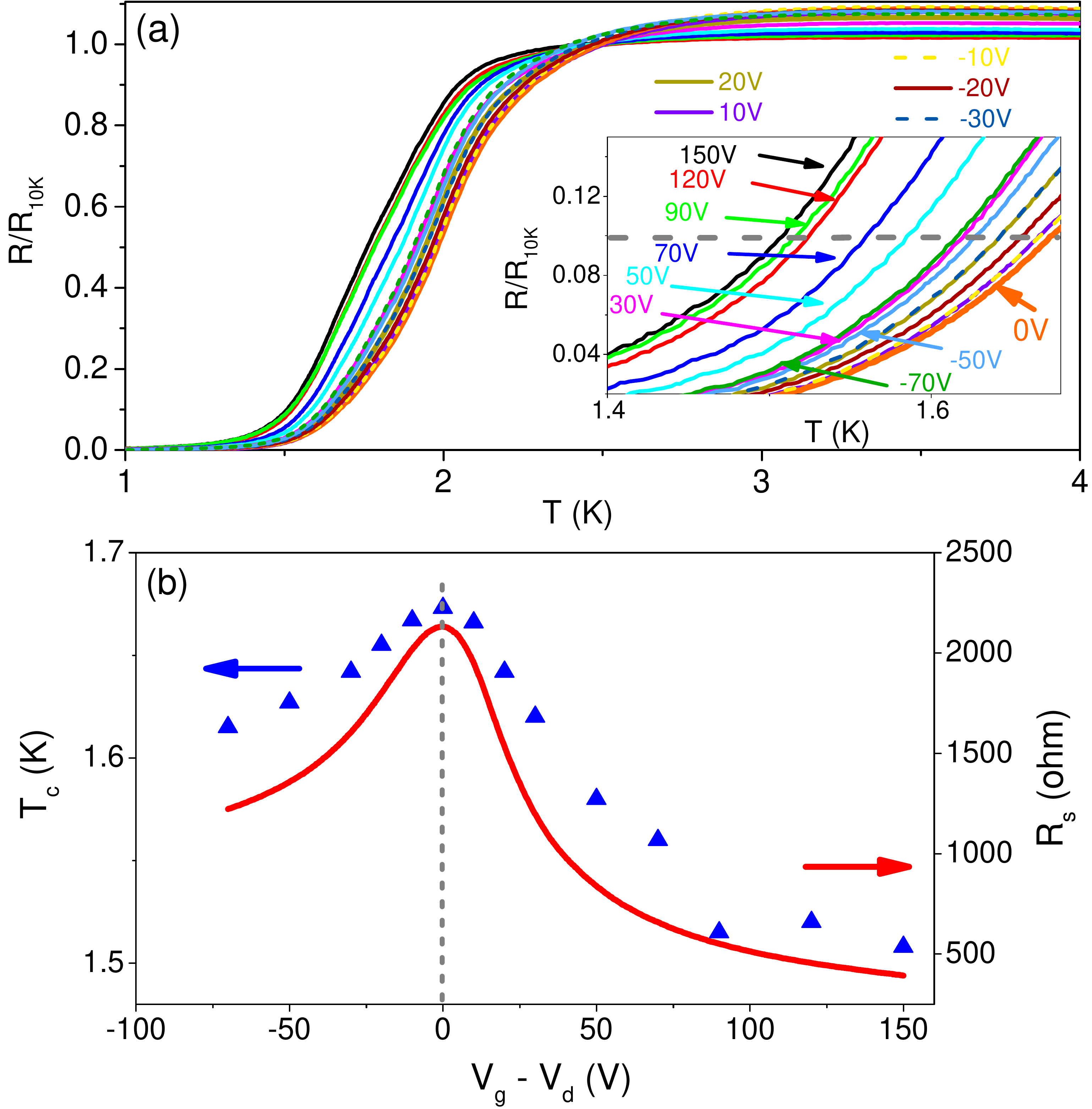}
\small{\caption{(a) Sheet resistance, $R_s$, normalized by the resistance at 10K, as a function of temperature at different gate voltages relative to the CNP, $V_g- V_d$, of sample Gr/S.  Inset: A zoom on the small temperature range highlighting the evaluation of $T_c$ with gate voltage. (b) $T_c$ and $R_s$ at $T=5$K as a function of $V_g-V_d$ measured at $B=0$T.
\label{RT}}}
\end{center}
\end{figure}

This notion is further supported by the R(T) curves at different $V_g$ presented in Fig.~\ref{RT} for sample Gr/S. For all gate voltages 
the heterostructure shows superconductivity at low temperatures. However, it is seen that $T_C$ (defined as the temperature at which 
the resistance drops to $90\%$ of the normal sheet resistance at 10K) systematically increases as $n$ decreases and reaches a 
maximum around the high-temperature CNP. This is in stark contrast with the common behavior of conventional superconductors and 
with previous experiments of Sn dots on graphene~\cite{allain2012electrical} which exhibit a minimum of $T_C$ at the CNP.  

A possible explanation for such behavior would be to invoke a non-BCS pairing mechanism in the proximitized islands in graphene. 
Such mechanisms have been used to explain the enhancement of superconductivity of exotic low density superconductors~\cite{PhysRevLett.114.077001}. 
However there seems to be no reason to assume that superconductivity in InO is of unconventional nature and hence any superconducting 
regions in the proximitized graphene are unlikely to show non-BCS properties.
Instead we suggest that in our samples, the graphene provides a medium for Josephson coupling between the superconducting clusters of InO, 
thereby enhancing the superfluid stiffness. We emphasize that, just as in the bare InO thin films, $T_{C}$ is dictated by the stiffness which controls phase fluctuations among the superconducting clusters, and not by the pairing amplitude. In the Gr/S heterostructure,
it is maximal close to the average CNP because 
(as shown below) Josephson effect through puddles of opposite polarity in the graphene layer is strongest when their average density is close to zero.

In clear contrast to sample Gr/I, in sample Gr/S the volume fraction of superconducting islands within the InO is roughly equal to that
of the insulating regions (see Fig.~\ref{sketch}c). In the underlying SLG, this generates large hole-doped puddles with proximity-induced superconductivity embedded in an electron-doped background. 
These superconducting islands are absent at temperatures far above $T_C$ since no emergent granularity is expected in the 
normal state~\cite{bouadim2011single} of InO. In this case, both electron-doped and hole-doped regions in the SLG contribute 
equally to the transport. Thus, Hall measurements feature a CNP (consistent with the observation of a peak in the (longitudinal) 
resistance as a function of the gate voltage [Fig.~\ref{RVg}(d)] for $T$ larger than 2.5K) when the average density 
of the sample is zero, i.e. the electron and hole densities are roughly equal. 
However, as $T$ is reduced and transport flows mostly through the superconducting islands, the finite resistance is dominated by patches of the SLG underlying the narrow constrictions between them. These effectively become SNS junctions where the S regions are hole-doped compared to the N region.

The proper model for the system at low $T$ is therefore 
a random array of Josephson junctions, where the Josephson coupling (ultimately dictating $T_C$ of the network) is provided by SNS constrictions of varying sizes. To analyse their $V_g$-dependence, we consider a single
SNS weak link 
and calculate its critical current ($I_c$) at $T=0$ (see Supplementary Material for details). The Fermi energy in the normal (N) region ($E_{F}$) is assumed to be positive, while the Fermi energy in the superconducting (S) regions  ($E_{F}^{\prime} = E_{F} - U$) is negative. The difference between the two ($U$) is assumed to arise from the difference in the electrostatic potential induced by the superconducting puddles in the InO. When $V_g$ is varied, $E_{F}$ and $E_{F}^{\prime}$ shift while maintaining $U$ fixed. The Josephson 
coupling of the junction is proportional to its critical current $I_{c}$. 
The length ($L$) of the weak link is assumed to be much smaller than the superconducting coherence length ($\xi$). 
In this limit, the contribution to the supercurrent from the continuum states may be neglected, and only the contribution from the subgap ($\epsilon < \Delta$) Andreev bound states needs to be computed~\cite{BeenakkerPRL,Furusaki99}. 
We also assume $L \ll W$ (the width of the link) so that there is a single bound state for each transverse wave vector.

Graphene SNS junctions have been studied previously in great detail~\cite{PhysRevB.74.041401,Annica08,Balatsky16,Takane20}, including in the 
limit considered here~\cite{PhysRevB.74.041401}. However the previous works only considered the case where superconducting regions were heavily doped 
compared to the normal region. These studies find that $I_{c}$ is minimal at the Dirac point of the normal region and increases 
monotonically as  the carrier density $n$ is increased. This behavior is compatible with, e.g.,
the experiments based on granular Sn islands deposited on a single layer of 
graphene~\cite{allain2012electrical}, 
where the S regions are metallic superconductors.

Our model goes beyond previous works in that we relax the assumption of very heavily doped superconducting regions. Furthermore, in our case the unique scenario dictated by the experimental system forces   
us to explore the regime where the carrier densities in the superconductor and normal regions are close to each other in magnitude, but opposite in sign. We evaluate the spectrum of the subgap Andreev bound states $\epsilon_q^{\text{ABS}}$ in such a Josephson 
junction as a function of the phase difference between the superconducting regions ($\phi$). The equilibrium Josephson current may be
found through, 
\begin{equation}
I(\phi)=-4\frac{e}{\hbar}\sum_q \frac{\partial \epsilon_q^{ABS}}{\partial \phi}. 
\end{equation}
Here the factor of 4 accounts for the spin and valley degeneracies. The critical current $I_{c}$ is simply the maximal value of 
$I(\phi)$. As explained earlier, the behavior of $I_{c}$ as a function of the Fermi energy 
is expected to follow the variation of $T_C$ as a function of $V_{g}$ in the sample Gr/S.

\begin{figure}[t]
\begin{center}
\includegraphics[width=0.45\textwidth]{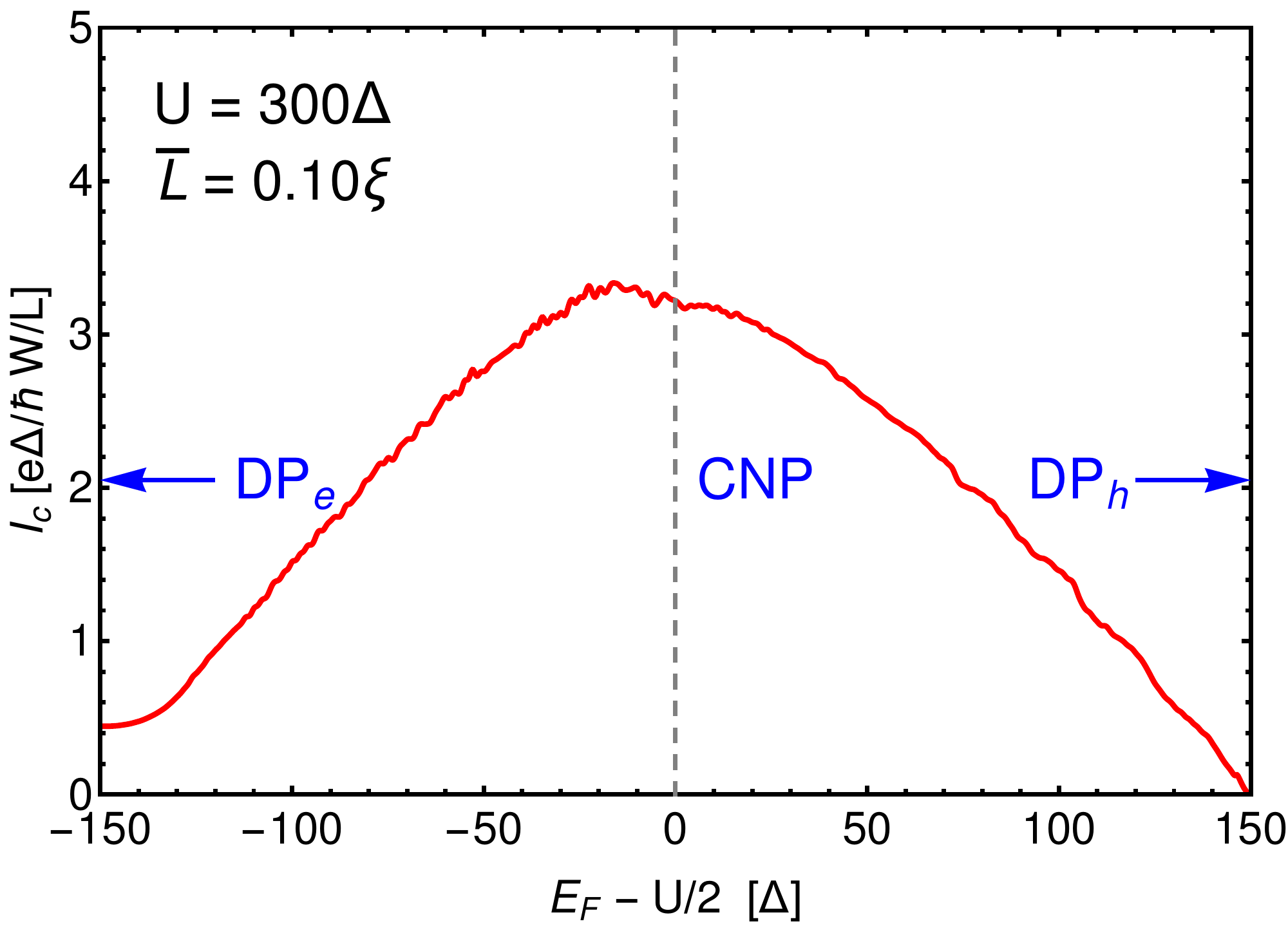}
\small{\caption{The critical current ($I_c$) as a function of the Fermi energy relative to the CNP (in units of $\Delta$). The leftmost (rightmost) energy corresponds to the DP$_e$ (DP$_{h}$). At the CNP, the carrier densities in the electron and hole regions are equal, so that the average density is zero. Contrary to the standard picture, $I_{c}$ is largest in the regime where the (net) carrier density is very small. This is a consequence of the opposite polarity of the superconducting and normal regions. The red curve shows $I_{c}$ after averaging over the length ($L$) of the SNS junction (keeping $\bar{L} = 0.1 \xi$), in order to remove the length-dependent features and account for the disordered nature of the superconducting puddles. Here, the value of the electrostatic shift is $U = 300 \Delta$. \label{fig:figure4}}}
\end{center}
\end{figure}

Fig.~\ref{fig:figure4} shows the variation of the critical current as a function of the Fermi energy ($E_{F}$) in the normal region. Note that in our convention, DP$_{e}$ (DP$_{h}$) appears at $E_{F} = 0$ ($E_{F}^{\prime} = E_{F} - U = 0$) which corresponds to the left (right) end of Fig.~4. The curve shown in Fig.~\ref{fig:figure4} was obtained after averaging $I_{c}$ over several 
values of $L$ (length of the SNS junction). The averaging removes spurious oscillatory features which depend on the value of 
$L$ (see SM), leaving behind a prominent gross feature: a broad maximum in the doping dependence of $I_{c}$. This captures the situation in 
the experimental system, where the percolating network of superconducting islands is expected to be dominated by several, most resistive, hotspots 
(or Josephson junctions) of varying lengths.

When the Fermi energy is close to the DP$_{e}$ 
our results match those reported in 
Ref.~\cite{PhysRevB.74.041401}, since the carrier density in the superconductors is quite large ($|E_{F}^\prime|\gg E_{F}$). With increasing $E_{F}$, 
the average $I_{c}$ increases monotonically until $E_{F} \sim U/2$. 
At this point, the carrier density in the S and N regions is equal, and the average carrier density of the SLG is 
expected to be close to zero. Hence, we expect $E_{F} = U/2$ to be close to the global CNP in our heterostructure (sample Gr/S). 
Increasing $E_{F}$ beyond $U/2$ drives the system into a previously unexplored regime, where the carrier density of the
normal region is larger than that of the superconductors. Andreev reflection at the two N-S interfaces is highly suppressed
in this regime, leading to a rapid decrease in $I_{c}$ despite the increase of carrier density in the normal region. 
For this reason, we observe the largest Josephson effect near $E_{F} = U/2$. Since the average CNP in sample Gr/S was
identified with this point, we expect to have the strongest Josephson coupling between the superconducting islands, 
and the largest enhancement in $T_{c}$, at the CNP. This is indeed consistent with our experimental observations (Fig.~\ref{RT}). 
Theoretically, $I_{c}$ has two local minimas at the Dirac point of the normal region ($E_{F} = 0$) and that of the superconducting
region ($E_{F} = U$). In our experiments however, we were unable to reach the two Dirac nodes, and only observed 
that the $T_C$ keeps decreasing away from the CNP.  

In summary, we have shown that coupling a SLG to a disordered, low-density superconductor leads to the unique result where superconductivity is strongest close to the average charge neutrality point of the graphene, in stark contrast to the situation in systems of SLG coupled to high-density superconductors. We ascribe this to the presence  of regions of opposite charge polarity induced within the graphene which acts as a coupling medium for superconducting islands. This newly explored regime provides access to Andreev reflections in low-density S-N junctions, where the carrier density in the superconducting regions is possibly lower than the normal ones. Furthermore, in the presence of magnetic field, the interplay between superconductivity in such heterostructure and the quantum Hall effect can give rise to intriguing phenomena. These will be the subject of future studies.  

\begin{acknowledgments}
We are grateful for help from I. Volotsenko, and useful discussions with J. Ruhman and N. Trivedi. G.N.D., A.R and A.F were supported by the Israel Science fund, ISF, grant No. 1499/21 and the US-Israel Binational Science Foundation (BSF) grant No. 2020331. U.K., E.W. and E.S. were supported by Israel Science Foundations (ISF) grant No. 993/19, the
US-Israel Binational Science Foundation (BSF) grant No.
2016130, and NSF-BSF grant No. 2018726. 
\end{acknowledgments}





%


\pagebreak

\vspace{1cm}

\section
{Supplementary information - Enhancement of Superconductivity upon reduction of carrier density in proximitized graphene}

	The Supplemental Material presents additional experimental data (section I, II, III) and details regarding the theoretical model (section IV).

\section{I. \,\, Hall measurement in single layer graphene}

\begin{figure}[!t]
	\includegraphics[width=0.98\columnwidth]{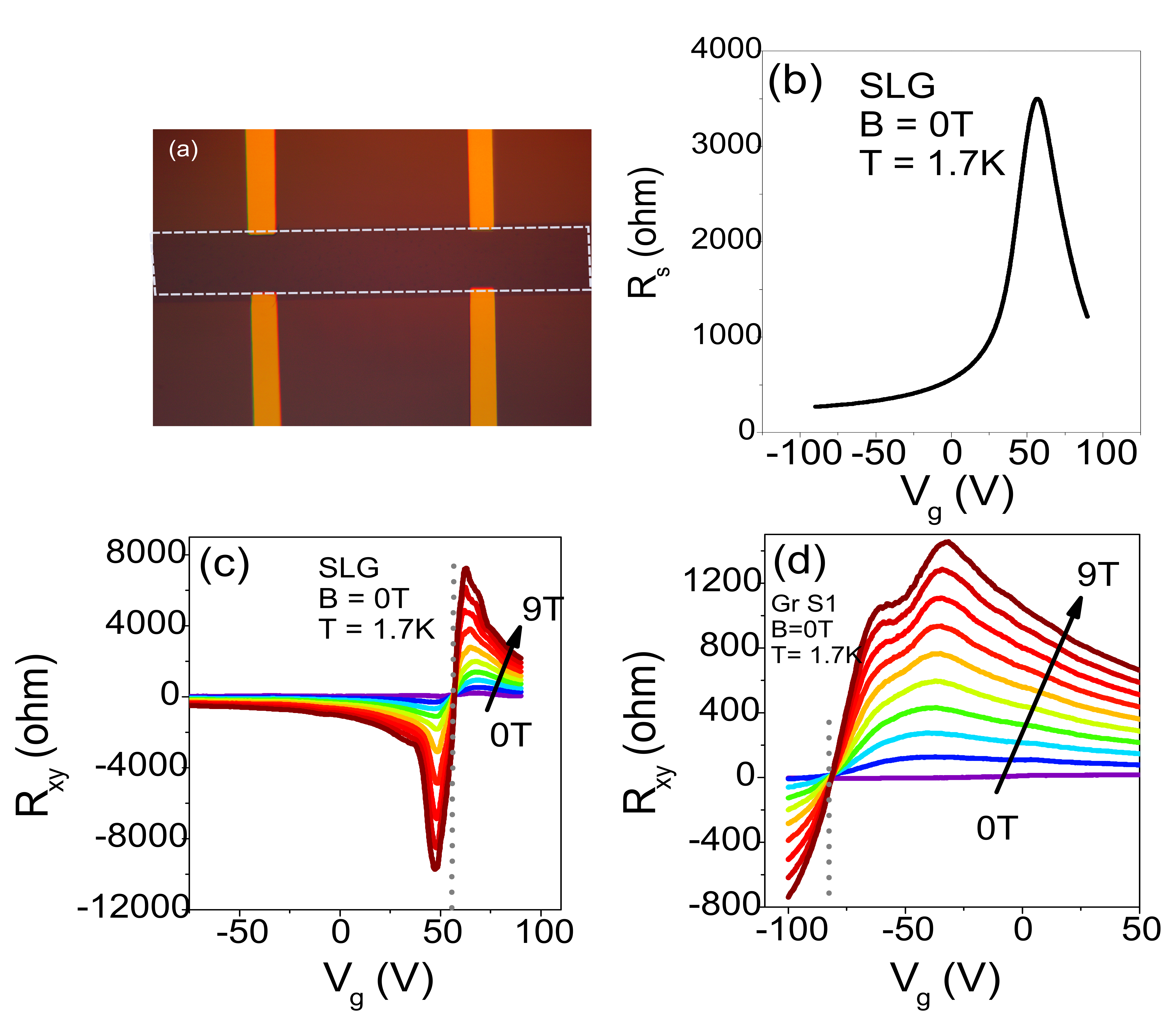}
	\small{\caption{ (a) Optical image of the CVD grown single layer graphene (see dashed white line). (b) Sheet resistance, $R_s$ of SLG as a function of gate voltage, $V_g$ at $B=0$ T and $T=1.7$ K. (c) and (d) Hall resistance $R_{xy}$ as a function of $V_g$ of SLG and sample Gr/S at different $B$ (in step of 1T) at $T$= 1.7~K. \label{SLG_RVg}}}
\end{figure}

Polycrystalline single layer graphene were prepared by the CVD technique on copper catalyst and were then transferred to 
SiO$_2$/Si substrate. The CVD grown single layer graphene samples were purchased from Graphenea Company which provided Raman data that shows that the entire graphene is single-layer. In Fig. \ref{SLG_RVg} (a), we show the optical image of CVD grown SLG.
Figure~\ref{SLG_RVg}(b) shows the sheet resistance as a function of $V_g$ at $B=0$ T and $T=1.7$ K. We identified the Dirac point (DP) as the charge neutrality point (CNP) extracted from the Hall measurement. In Fig.~\ref{SLG_RVg}(c), 
Hall resistance $R_{xy}$ is plotted as a function of $V_g$ for different magnetic field and $T = 1.7$ K. Note that the DP is at $V_d=56$ V. This indicates that graphene is hole doped due to adsorption of atmospheric dopants such as H$_2$O and O$_2$ \cite{shin2010surface}. When the graphene is coated by InO films having increasing electron carrier density, incresasing electron doping is induced in the graphene thus shifting the charge neutrality point to negative gate voltages, until, for our superconducting sample Gr/S the CNP reaches ~-81.5V (see Fig. \ref{SLG_RVg}(d)).

\begin{figure}[!t]
	\includegraphics[width=0.98\columnwidth]{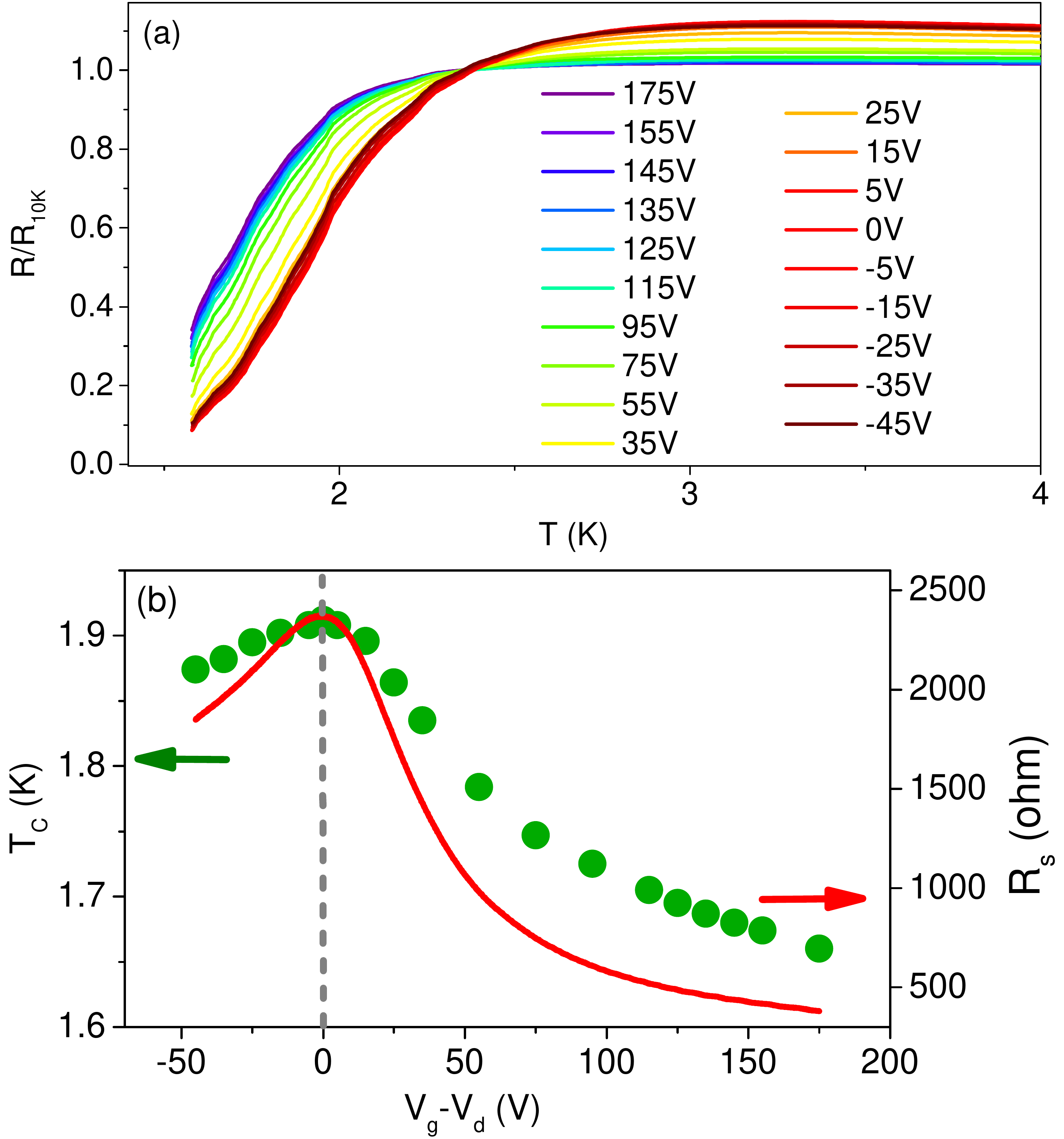}
	\small{\caption{ Sheet resistance, $R_s$, normalized by the resistance at 10K, as a function of temperature at different gate voltages relative to the CNP, $V_g- V_d$, of sample Gr/S2. (b) $T_c$ and $R_s$ at $T=5$K as a function of $V_g-V_d$ measured at $B=0$T. \label{SI_RT}}}
\end{figure}

\section{II. \,\, Resistance-temperature curves at different gate voltage for sample Gr/S2}
Figure \ref{SI_RT}(a) shows the sheet resistance, normalized by the resistance at 10K, as a function of temperature at different gate voltages relative to the charge neutrality point, CNP, $V_g-V_d$ for sample Gr/S2. The measurements were performed in a wet He-4 system at temperatures down to 1.5 K. For all gate voltages, sample Gr/S2 shows signature of superconductivity at low temperatures. Figure \ref{SI_RT}(b) shows that $T_C$ (defined as the temperature at which resistance drops to 50\% of the normal sheet resistance at 10 K) increases as $n$ decreases and becomes maximum close to the charge neutrality point.

\section{III. \,\, Hall resistance - gate voltages at different temperature}
\begin{figure}[!t]
	\includegraphics[width=0.7\columnwidth]{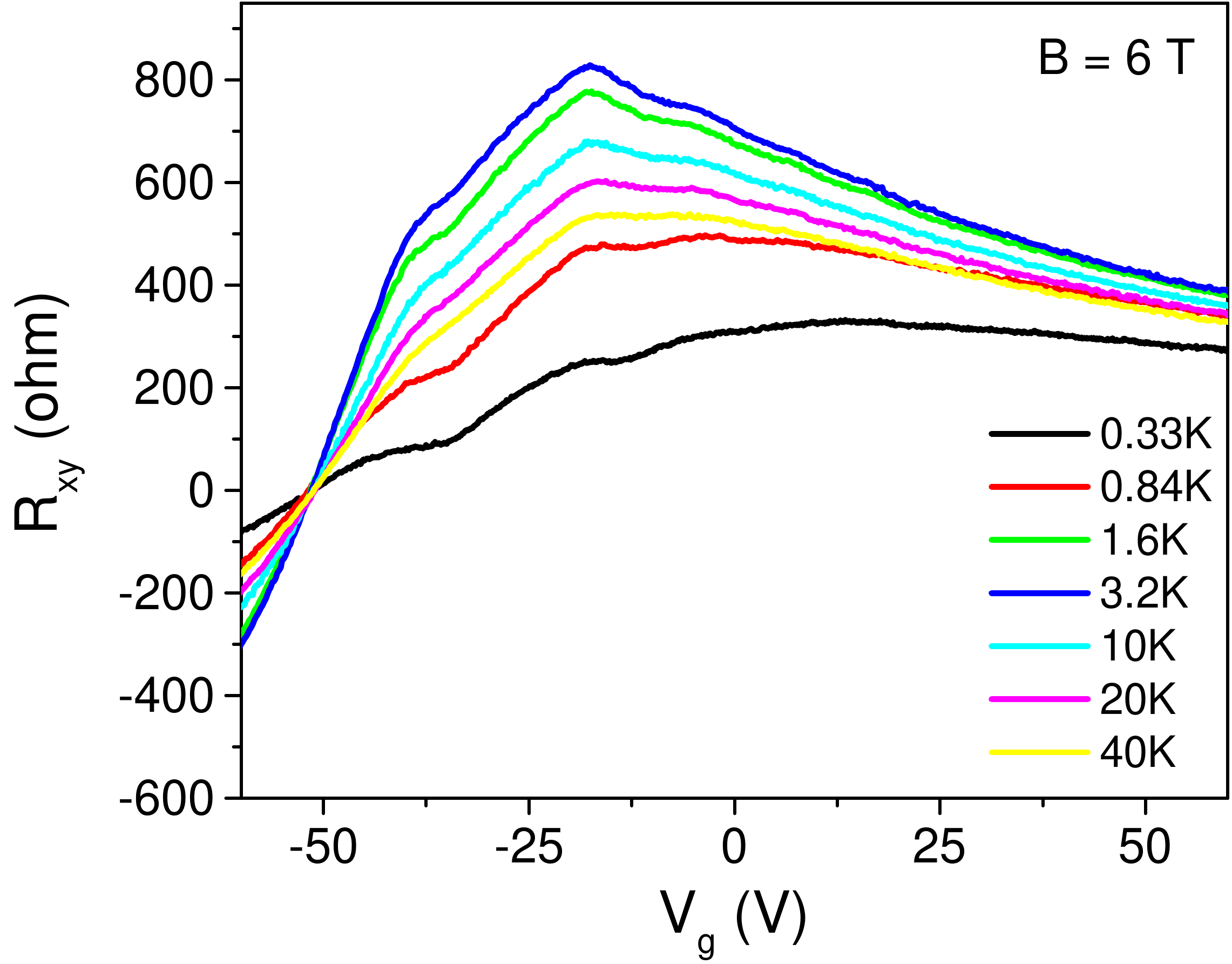}
	\small{\caption{ Hall resistance as a function of gate voltages at different temperatures of sample Gr/S2. The measurements were performed at $B=6$ T. \label{Hall_T}}}
\end{figure}

In Fig. \ref{Hall_T} , Hall resistance, $R_{xy}$, is plotted as a function of gate voltages at different temperatures for sample Gr/S2. The measurements were performed at $B =  6$ T. It can be seen that for temperatures below $T_c$, the Hall effect is suppressed until, for low enough temperatures it is expected to vanish.

\section{IV. \,\, The Model for a Single Junction}

\begin{figure}[t]
	\includegraphics[width=0.75\columnwidth]{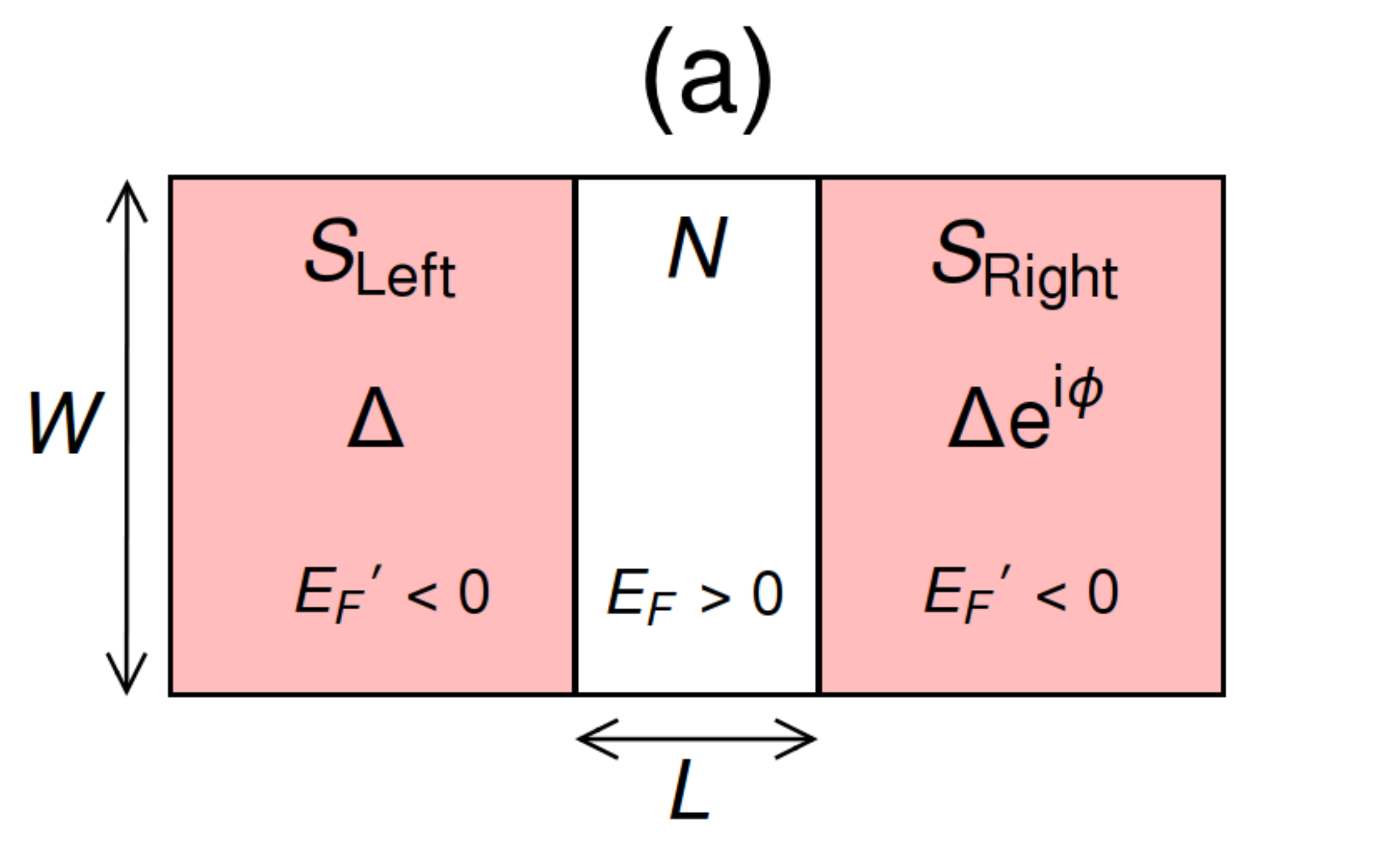}
	\includegraphics[width=0.98\columnwidth]{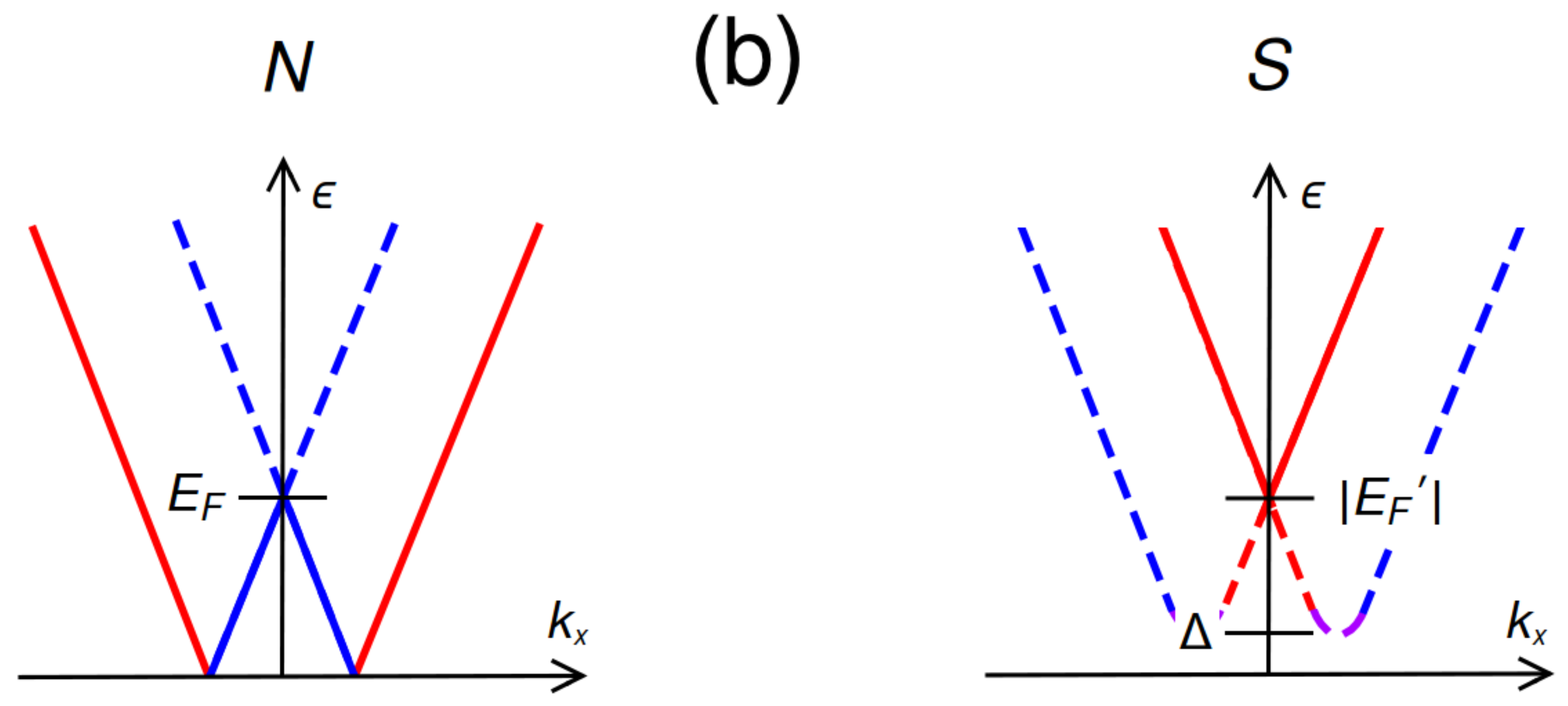}
	\small{\caption{ (a) Schematic of the SNS junction studied here. (b) Low-energy excitations in the normal (N) and superconducting (S)
			regions (at $q = 0$) of the junction. The solid (dashed) lines denote states corresponding to conduction (valence) bands, while the 
			red (blue) color denote the electron-like (hole-like) states in the Nambu representation. Note that the N (S) region is considered to be
			electron-doped (hole-doped), so that $E_{F} > 0$ ($E_{F}^{\prime} = E_{F} - U < 0$). Therefore the excitations with largest $k_{x}$, 
			are electron-like states of the conduction band (hole-like states of the valence band) in N (S) regions. 
			\label{SAB}}}
\end{figure}

We consider a superconductor-normal-superconductor (SNS) junction of length $L$ (along the $x$ direction) in the ballistic limit [Fig.~\ref{SAB}(a)]. Here, the superconducting electrodes correspond to  proximitized regions of graphene, with a proximity-induced pairing potential $\Delta$.  
Our aim is to evaluate the critical 
current ($I_{c}$) in the junction as a function of the gate voltage. We assume that $L \ll \xi$ (the superconducting coherence length). 
In this limit, the Josephson current is carried mostly by the Andreev bound states (ABS) with energies ($\epsilon^{\text{ABS}}_{q}$) 
smaller than $\Delta$. Here $q$ could be any generic quantum number labelling the ABS. We consider 
N-S interfaces parallel to the $y$ direction, and 
impose periodic boundary conditions along $y$. We further assume that the width ($W$) of the junction is much larger than its length $L$. 
Under these approximations, $q = 2\pi n/ W$ (for $n \in \mathbb{Z}$) denotes the wave vector along $y$, and there is a single ABS for each $q$. 
We stress that we consider these limits only to simplify the calculation of $I_{c}$. We do not expect our results to change qualitatively for 
longer Josephson junctions. 

This geometry was also considered by Titov and Beenakker (Ref.~[22]). However, unlike previous 
works, we consider the case where the superconducting and normal regions have opposite polarities. Furthermore, we do not assume that 
the Fermi energy in the superconducting puddles ($E_{F}^{\prime}$) is much larger than the Fermi energy of the normal regions ($E_{F}$). 
Our analysis relies on completely general solutions of the BdG equations in the superconducting regions. By contrast, the wave functions
employed in Ref.~[22] were limiting cases of Eqs.~(\ref{eq:PsiSA}-\ref{eq:PsiSB}) below.  

For a fixed phase difference ($\phi$) between the superconductors, the junction supports an equilibrium supercurrent, 
$I(\phi)$, which depends on the ABS spectrum through, 
\begin{align} \label{eq:Ip}
I(\phi) = -4 \frac{e}{\hbar} \sum_{q} \frac{\partial \epsilon^{\text{ABS}}_{q}}{\partial \phi}. 
\end{align}
The factor of 4 accounts for the spin and valley degeneracies. 
The critical current, $I_{c}$, is the largest value of $I(\phi)$. 
Therefore our main task is to evaluate the energy of subgap ABS as a function of $q$ and $\phi$. 

Due to Andreev reflection at the N--S interfaces, a general state inside the normal region of graphene is a superposition 
of electrons and holes from different valleys. For brevity, we only consider states involving electrons from a given valley and with 
a fixed spin polarization. The spin-valley degeneracy is incorporated in the end through the factor of $4$ in Eq.~(\ref{eq:Ip}).
Thus the dynamics of the normal region may be described through the Hamiltonian,  
\begin{align} \label{eq:HN}
H_N = \left( \begin{array}{cc}
H_0 - E_F & 0 \\
0 & E_F - H_0 \end{array} \right), 
\end{align}
where $H_0 = -i (\sigma_x \partial_x + \sigma_y \partial_y)$, and $\sigma$ represents 
the sublattice degree of freedom. Throughout our analysis $\hbar v_F$ is set to 1, and we employ the Nambu representation. 
The upper (lower) block of $H_{N}$ corresponds to electrons (holes). The Hamiltonian for electrons
in the other valley is similar to $H_{N}$ with $H_{0} \rightarrow -i (\sigma_x \partial_x - \sigma_y \partial_y)$. Eq.~(\ref{eq:HN}) 
has 4 eigenvectors for a given energy ($\epsilon$) and transverse wave vector ($q$), corresponding to the left and right moving 
electrons ($\psi_{e,L}$ and $\psi_{e,R}$) and holes ($\psi_{h,L}$ and $\psi_{h,R}$). Consequently a general state in the normal
region with a given $\epsilon$ and $q$ is,
\begin{align}
\Psi_{N} (\epsilon, q) = a_{+} \psi_{e, R} + a_{-} \psi_{e, L} + b_{+} \psi_{h, R} + b_{-} \psi_{h, L}. 
\end{align}
Note that the dependence on position has been suppressed for brevity. 

The proximitized regions of graphene are described by the Hamiltonian,
\begin{align} \label{eq:HS}
H_S = \left( \begin{array}{cc}
H_0 - E_{F}^{\prime} & \Delta e^{i \phi} \\
\Delta e^{-i \phi} & E_{F}^{\prime} - H_0 \end{array} \right)
\end{align}
where, $E_{F}^{\prime}$ is the Fermi energy in the superconducting region.
$E_{F}^{\prime}$ is positive (negative) in electron-doped (hole-doped) superconductors. For an interface between a  metallic superconductor (such as a grain of Sn) and graphene,
$E_{F}^{\prime}$ is not only positive but $E_{F}^{\prime} \gg E_{F}$. This is the limit considered in Ref.~[22]. In our case, 
$E_{F}^{\prime} = E_{F} - U$ is assumed to be negative. We assume that the electrostatic shift ($U$) is fixed by the coupling between graphene and InO, while $E_{F}$ (and consequently $E_{F}^{\prime}$) may be tuned using external gates.  

\begin{figure}[t]
	\includegraphics[width=0.98\columnwidth]{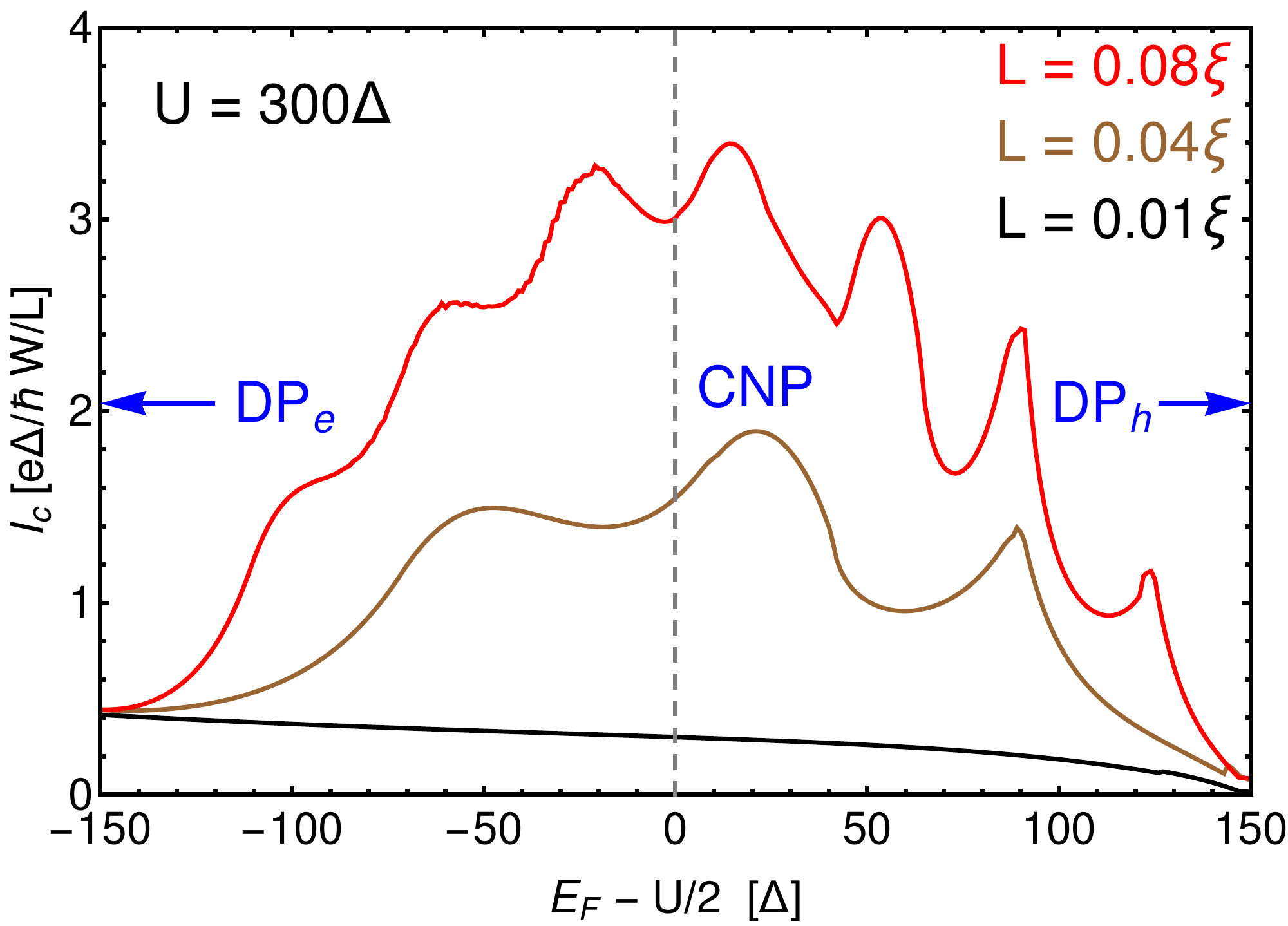}
	\small{\caption{ The critical current ($I_c$) as a function of the Fermi energy relative to the CNP (in units of $\Delta$). 
			The different curves denote $I_{c}$ for junctions of different lengths. The leftmost (rightmost) energy corresponds to the 
			DP$_e$ (DP$_{h}$). At the CNP, the carrier densities in the electron and hole regions are equal, so that the average density is zero. 
			Note that for sufficiently large $L$, $I_{c}$ has an overall maxima close to the CNP, along with several length-dependent 
			oscillations which arise from the Andreev bound states. These length-dependent features would only be observable in very clean
			devices comprising a single Josephson junction. In the disordered samples considered here, we only observe the envelope function
			of $I_{c}$ which is maximal close to the CNP. 
			\label{SIc}}}
\end{figure}

$H_{S}$ has four eigenvectors for a given $\epsilon$ and $q$. For $\epsilon < \Delta$ the wave vector along $x$: $p_{x}$, is complex. 
Hence, inside a given superconductor only the two solutions which decay with distance away from the interface are physically relevant for our setup. 
We label the physical solutions as $\psi_{\pm, L/R}$, where the sign $\pm$ labels the two solutions in the superconductor to 
the left ($L$) or right ($R$) of the normal region. Solving the eigenvalue equation corresponding to $H_{S}$, we find, 
\begin{align} \label{eq:PsiSA}
\psi_{+, L/R} &= 
e^{i (k_{x} x + q y) - \zeta \kappa x} \left( 
\begin{array}{c}
e^{i \zeta \times \text{sign}(E_{F}^{\prime}) \beta} \\ 
d_{+} e^{i \zeta \times \text{sign}(E_{F}^{\prime}) \beta} \\ e^{-i \phi} \\ d_{+} e^{-i \phi}
\end{array} \right), 
\end{align}
\begin{align} \label{eq:PsiSB}
\psi_{-, L/R} &= 
e^{i (- k_{x} x + q y) - \zeta \kappa x} \left( 
\begin{array}{c}
e^{-i \zeta \times \text{sign}(E_{F}^{\prime}) \beta} \\ 
d_{-} e^{-i \zeta \times \text{sign}(E_{F}^{\prime}) \beta} \\ 
e^{-i \phi} \\ d_{-} e^{-i \phi}
\end{array} \right). 
\end{align} 
Here we have used $\zeta = +1 (-1)$ for the superconductor on the right (left), $\beta = \arccos[\epsilon / \Delta]$, 
$k_{x} = |\text{Re}[p_{x}]|$, and $\kappa = |\text{Im}[p_{x}]|$. Finally, 
\begin{align}
d_{\pm} &=  
\frac{E_{F}^{\prime} \pm i \zeta \times \text{sign}(E_{F}^{\prime}) \sqrt{\Delta^{2} - \epsilon^{2}}}{p_{x \pm} - i q}.
\end{align} 
Thus, the most general wave function (given $\epsilon$ and $q$) in the two superconducting regions is, 
\begin{align}
\Psi_{S, L/R}(\epsilon, q) = c_{+, L/R} \psi_{+, L/R} + c_{-, L/R} \psi_{-, L/R}. 
\end{align} 
As an aside we note that, in the limit $E_{F}^{\prime} \gg E_{F},\Delta > 0$, 
\begin{align}
d_{\pm} \approx \pm e^{\pm i \gamma} \,\, \text{ where } \gamma = \sin^{-1}
\left( \frac{q}{|E_{F}^{\prime}|} \right). 
\end{align}
In this limit, the solutions above [Eqs.~(\ref{eq:PsiSA}-\ref{eq:PsiSB})] reduce to those given in Ref.~[22].  

To solve for the ABS we impose continuity conditions on the wave functions, $\Psi_{S}$ and $\Psi_{N}$, at the two N--S interfaces. 
For a given $q$, the boundary conditions $\Psi_{S L} (x = -L/2) = \Psi_{N} (x = -L/2)$ and $\Psi_{S R} (x = L/2) = \Psi_{N} (x = L/2)$ 
can be satisfied simultaneously only for some discrete values of $\epsilon < \Delta$. These are the ABS localized in the normal region. 
We find the ABS spectrum $\epsilon_{q}^{\text{ABS}}$ by numerically solving the equations for the coefficients 
($a_{\pm}, b_{\pm}, c_{\pm, L/R}$) which arise from the boundary conditions. 
The current is then found by replacing the derivative $\partial \epsilon / \partial \phi$ [in Eq.~(\ref{eq:Ip})] with 
a (central) finite difference expression.

Fig.~\ref{SIc} 
shows the variation of $I_{c}$ with $E_{F} - U/2$ for different values of $L$. Note that in our notation, $E_{F} - U/2 = -U/2$ corresponds to DP$_{e}$ (the Dirac point in normal regions of graphene, defined by $E_{F} = 0$), while $E_{F} - U/2 = U/2$ corresponds to DP$_{h}$ (the Dirac point in superconducting regions of graphene, defined by $E_{F}^{\prime} = 0$). In between these values, $E_{F} > 0$ while $E_{F}^{\prime} < 0$. Our results 
clearly show that while several details of the $I_{c}$ vs. $E_{F}$ curve depend on the length $L$ of the junction, the critical current (for $\frac{U}{\hbar v_{F}} L \gg 1$) is maximal close to the average CNP, defined as $E_{F} = U/2=E_{F}^{\prime}$. Such $L$-dependent features would be suppressed in an array of several junctions with a random variation of lengths. Therefore Fig.~4 of the main text depicts $I_{c}$ averaged over several values of $L$ (keeping $\bar{L} \ll \xi$). We stress that the average $I_{c}$ should be maximal close to the CNP, but since there is no symmetry relating the DP${_e}$ and DP$_{h}$, there is no reason to expect that maximal value to be exactly at the CNP. Furthermore although the averaging procedure strongly suppresses the oscillations, some small wiggles would be left behind (which are manifest in Fig.~4 of the main text).

\end{document}